\documentclass[amsmath,superscriptaddress,
amssymb,aps,prd,floatfix,twocolumn]{revtex4}

\usepackage{xcolor}
\usepackage{graphicx}
\usepackage{mathrsfs}
\usepackage[bookmarks=false]{hyperref}
\usepackage[utf8]{inputenc}
\usepackage{dcolumn}
\usepackage{bm}
\usepackage{soul}

\begin{document}

\title{Phase diagram and compact stars in a holographic QCD model}

\author{Luis A. H. Mamani}%
\email{luis.mamani@ufabc.edu.br}
\affiliation{
Centro de Ci\^encias Naturais e Humanas, Universidade Federal do ABC, Avenida dos Estados 5001, 09210-580 Santo Andr\'e, São Paulo, Brazil}

\author{Cesar V. Flores}
\email{cesar.vasquez@uemasul.edu.br}
\affiliation {Centro de Ci\^encias Exatas, Naturais e Tecnológicas, UEMASUL,
Rua Godofredo Viana 1300, Centro
CEP: 65901- 480, Imperatriz, Maranhão, Brazil}

\author {Vilson T. Zanchin}
\email{zanchin@ufabc.edu.br}
\affiliation {
Centro de Ci\^encias Naturais e Humanas, Universidade Federal do ABC, Avenida dos Estados 5001, 09210-580 Santo Andr\'e, São Paulo, Brazil}

\begin{abstract}

A holographic model is used to investigate the thermodynamics and the phase diagram of a heavy quarks system. From such a model we obtain an equation of state and explore its applicability in astrophysical conditions. For this objective, we work in the context of the Einstein-Maxwell-Dilaton (EMD) holographic model for quantum chromodynamics (QCD). At first, we show the existence of a critical point where the first-order transitions line ends, later on, we calculated an analytic expression for the equation of state. Additionally,  with the aim of investigating the global properties of compact stars, such as the total gravitational mass and radius, the equation of state is used to solve the Tolman-Oppenheimer-Volkov (TOV) equations for stellar structure. The numerical results show that our equation of state is able to reproduce the expected behavior of hybrid stars. Our main conclusion is that, by using an equation of state emerging in the framework of the EMD holographic model for QCD, it is possible to obtain quark matter properties and that it is also possible to extend the procedure to astrophysical applications.
\end{abstract}

\maketitle

\section{Introduction}

The investigation of the phase structure of quantum chromodynamics (QCD) is an open problem of modern physics, and there are several research groups around the world facing this problem. It is widely known that QCD lies in the confinement regime in the region of low temperature $T$ and density (chemical potential $\mu$) and that it also lies in the deconfinement regime in the region of high temperature and density. It is believed that at the boundary between these two phases, close to the chemical potential axis, there is a line describing first-order phase transitions. It is also speculated that this line terminates at the critical point, where the theory has conformal symmetry and can be described by a set of universal critical exponents. In turn, at low chemical potentials, the transition becomes crossover. Moreover, it is complicated to extract reliable information from the region where these transitions occur because QCD lies in the strong coupling regime where the usual techniques used in perturbative QCD do not work. 
On the other hand, it is known that lattice QCD provides reliable results at zero chemical potential (in the strong coupling regime), however, it does not work when finite chemical potential effects are considered. Among other problems arising at finite density is the sign problem. Nevertheless, most of the relevant problems of modern physics are related to QCD at finite density, for example, heavy-ion collisions and investigation of compact objects in astrophysics. In the last years, lattice QCD is overcoming the application problems by implementing new techniques, but these do work in the regime of small chemical potentials and no reliable results at large densities are available so far, see e.g.~\cite{Philipsen:2010gj, Karsch:2007dp, Satz:2008kb} for a review. 

Another theoretical framework for investigating QCD-like theories at finite temperature and density has been implemented following the holographic principle \cite{Stephens:1993an, Susskind:1994vu}. The anti-de Sitter/conformal field theory (AdS/CFT) correspondence \cite{Maldacena:1997re, Gubser:1998bc, Witten:1998qj}, which is a realization of the holographic principle, allows us to investigate the duality between a strongly coupled field theory living in a $d$-dimensional spacetime and its dual gravitational theory (in principle quantum gravity) living in a $(d+1)$-dimensional spacetime. The duality arising directly from superstring theory in ten or eleven spacetime dimensions is known in the literature as a top-down approach, see for instance Ref.~\cite{Karch:2002sh, Sakai:2004cn, Sakai:2005yt, Misra:2019thm, Czajka:2018bod}. On the other hand, the bottom-up approach relates quantum field theories living in four-dimensional spacetimes with a dual classical gravitational theory living on five-dimensional anti-de Sitter spacetime. These models have phenomenological motivations and are build using phenomenological results in the quantum field theory side. 
In the first stages the backreaction on the gravitational background was neglected \cite{Erlich:2005qh, Karch:2006pv, Brodsky:2003px} (see also \cite{Miranda:2009uw, Mamani:2013ssa, Braga:2016wkm, Braga:2017fsb, Mamani:2018uxf, Braga:2019jqg, Braga:2020myi}), then, the backreaction is considered \cite{Csaki:2006ji, Gursoy:2007cb, Gursoy:2007er, Batell:2008zm, dePaula:2008fp, Li:2013oda, Ballon-Bayona:2017sxa, Mamani:2019mgu} (see also \cite{Dudal:2017max, Bartz:2018nzn, Chen:2019rez, FolcoCapossoli:2019imm, MarinhoRodrigues:2020ssq}), this approach is also known as holographic QCD. In the context of holographic QCD models it was shown that scenarios with simple gravitational theories in five-dimensional spacetime are able to mimic certain properties of QCD, like the equation of state (EoS) \cite{Gubser:2008ny, Gubser:2008yx} and the thermodynamics of the gluon plasma \cite{Gursoy:2008bu, Gursoy:2008za, Gursoy:2009jd, Gursoy:2009kk, Noronha:2009ud}. What is interesting is that these results were obtained with a  gravitational action coupled to a scalar field. The conformal symmetry breaking is realized by the nontrivial profile of the scalar field. In the same way, by adding an additional U(1) gauge field in the gravitational action we may include finite chemical potential in the dual field theory. Thus, the gauge field in the five-dimensional theory is dual to the baryon density current and it may be generated turning on an electric field in the gravitational background. It is worth mentioning that the resulting metric is asymptotically Reissner-Nordtr\"om AdS metric.

Previous works investigated the phase structure and critical point using  holographic QCD at finite temperature and density \cite{DeWolfe:2010he, DeWolfe:2011ts, Cai:2012xh, He:2013qq, Yang:2014bqa, Finazzo:2016psx, Knaute:2017opk, Li:2017ple, Sin:2007ze, Colangelo:2010pe, Ballon-Bayona:2020xls}. It is worth pointing out that in \cite{DeWolfe:2010he, DeWolfe:2011ts} the authors found a phase diagram in agreement with that is expected in QCD, they also found a critical point. Interestingly, the information concerning the quarks seems to be codified in the gauge field and a scalar function, gauge kinetic function, which characterizes the non-minimal coupling between the dilaton and gauge field. It is also interesting to point out that the authors of Ref.~\cite{He:2013qq} investigated a holographic model for heavy quarks, which phase diagram differs from the one obtained in \cite{DeWolfe:2010he, DeWolfe:2011ts}. Phenomenological properties, like mesons dissociation, were also investigated using holographic models, see for instance Refs.~\cite{Lee:2009bya, Park:2009nb, Park:2011qq, Colangelo:2011sr}.

On the other hand, the interest in investigating neutron stars increased promoted by the results collected by the LIGO and VIRGO collaborations \cite{TheLIGOScientific:2017qsa}. Previous investigations of compact objects, like neutron stars, were developed using the Tolman \cite{Tolman:1939} and Oppenheimer-Volkov \cite{Oppenheimer:1939} equations. A pivotal point to solve these equations is the equation of state (EoS) of matter in the interior of these objects, where it is under extreme conditions. The results obtained applying usual methods in perturbative QCD, for example, are not reliable because of the high density and strong coupling. It is also believed that the collision of neutron stars will constraint, even more, the EoS in such extreme conditions. At this point, the holographic approach seems to be useful for investigating the nuclear matter in such conditions. As mentioned above, this theoretical framework maps the problem into a dual classical gravitational theory, where, in principle, we can face the problem using techniques available in the literature. This way to face the problem will allow us to shed new light in the understanding of compact objects, like neutron, quark matter, or even hybrid stars.

In this paper, we work with the holographic model proposed in Ref.~\cite{Gursoy:2007er} (see in particular Appendix G of such Ref.), then applied it to investigate Yang-Mills theories at finite temperature in Ref.~\cite{Caselle:2011mn}. To include finite chemical potential effects in the dual field theory we add an abelian gauge field in the five-dimensional action following Refs.~\cite{DeWolfe:2010he, DeWolfe:2011ts}. Thus, the holographic model describes the heavy quarks system in the dual field theory \cite{He:2013qq}. One of the aims of this work is to apply our EoS to solve the Tolman-Oppenheimer-Volkov-(TOV) equations in order to investigate the behavior of the mass vs radius relationship for hybrid stars. As far as we know previous works in the literature have investigated the internal structure of stars using holography, see for instance Refs.~\cite{Fadafa:2019euu, Ecker:2019xrw, Chesler:2019osn, Hirayama:2019vod, Jokela:2018ers, Hoyos:2016zke, Annala:2017tqz, Hoyos:2020hmq, Jokela:2020piw}. The advantage of working with the model of Ref.~\cite{Caselle:2011mn} is that we get analytic solutions for some of the relevant thermodynamic variables like temperature and entropy.

The paper is organized as follows. In Section \ref{Sec:5DimensionalModel} we introduce the five-dimensional model. We also solve the Einstein-Maxwell-Dilaton equations using a simple holographic model and get an analytic expression for the gauge field, temperature, and entropy density. The asymptotic analysis of the thermodynamic variables allows us to get an analytic EoS. In the end, we investigate the extremal solution, i.e., when the temperature vanishes. In Section \ref{Sec:NumericalResults}, our numerical results are presented and discussed. Section \ref{Sec:PhaseDiagram} is devoted to investigating the phase structure of the holographic model. In Section \ref{Sec:TOV} we implement the matching procedure between the nuclear EoSs and the quark matter EoS, and then we solve the  Tolman-Oppenheimer-Volkov equations to get the mass-radius relationship. Finally, we conclude and discuss future extensions of the present work in Section \ref{Sec:Conclusions}. Complementary material is left in Appendix \ref{AppendixA}.

\section{The five-dimensional EMD model}
\label{Sec:5DimensionalModel}

\subsection{The five-dimensional background}
The gravitational dual of the Yang-Mills theory in four-dimensional spacetime is described by a five-dimensional gravitational theory whose action contains the metric coupled to a scalar field and, in order for including finite density effects in the dual field theory, a gauge field must be added in the five-dimensional action, and thus the metric is also coupled to this field. The most general five-dimensional action is defined by \cite{DeWolfe:2010he}
\begin{equation}\label{Eq:5DAction}
S=\frac{1}{2\kappa^2}\int dx^5\sqrt{-g}\left[R+\mathcal{L}\right],
\end{equation}
where the Lagrangian is given by
\begin{equation}
\mathcal{L}=-\frac{4}{3}(\partial^m\Phi)(\partial_m\Phi)
+V(\Phi)-\frac{f_{\scriptscriptstyle{\Phi}}(\Phi)}{4}F^{mn}F_{mn},
\end{equation}
where $\Phi$ is the dilaton field, $V(\Phi)$ is the dilaton potential and contains the cosmological constant term, $-6/\ell^2$, where $\ell$ is the AdS radius, $F_{mn}$ the gauge field defined by $F_{mn}=\partial_{m}A_n-\partial_{n}A_m$ and $\kappa$ is the Newton constant in five-dimensional spacetime, while $f_{\scriptscriptstyle{\Phi}}(\Phi)$ is the gauge kinetic function, which includes the non-minimal coupling between the dilaton and gauge field. In the present work we restrict ourselves to the case $f_{\scriptscriptstyle{\Phi}}(\Phi)=1$. The equations of motion obtained from varying the action \eqref{Eq:5DAction} are:
\begin{equation}\label{Eq:Einstein}
\begin{split}
&G_{mn}-\frac{4}{3}(\partial_m\Phi)(\partial_n\Phi)+\frac{2g_{mn}}{3}(\partial^p\Phi)(\partial_p\Phi)-\frac{g_{mn}}{2}V(\Phi)\\
&+\frac{1}{2}\left(\frac{g_{mn}}{4}F^{pq}F_{pq}-g^{pq}F_{nq}F_{mp}\right)=0,
\end{split}
\end{equation}
where $G_{mn}$ is the Einstein tensor. In turn, the Maxwell equations are:
\begin{equation}\label{Eq:Maxwell}
\begin{split}
\partial_m\left(\sqrt{-g}g^{mp}g^{nq}F_{pq}\right)=0.
\end{split}
\end{equation}
Finally, the Klein-Gordon equation
\begin{equation}
\frac{8}{3\sqrt{-g}}\partial_m\left(\sqrt{-g}\partial^{m}\Phi\right)+\partial_{\Phi}V(\Phi)=0.
\end{equation}
In applications of holographic QCD it is usual to consider an ansatz on the metric tensor and gauge field of the form \cite{Lee:2009bya,Park:2009nb}
\begin{equation}\label{Eq:AnsatzMetric}
\begin{split}
ds^2=e^{2{\cal A}(z)}\left(-f(z)dt^2+\frac{1}{f(z)}dz^{2}+dx_{i}dx^{i}\right),\\
A_0=A_0(z),\qquad A_1=A_2=A_3=A_4=0, 
\end{split}
\end{equation}
where ${\cal A}(z)$ is the warp factor, we also define the new function $\zeta=e^{-\cal A}$ to rewrite the equations of motion. Neglecting the dilaton field, the solution of the Einstein-Maxwell equations is the (fifth-dimensional) Reissner-Nordstr\"om AdS black hole (RNAdS). Following the convention of Refs.~\cite{Park:2009nb,Park:2011qq,Sin:2007ze}) we are going to consider the constant $\kappa$ given by
\begin{equation}\label{Eq:Relationskg5}
\frac{1}{\kappa^2}=\frac{N^2}{(2\pi)^2\ell^3},
\end{equation}
where $N$ is the number of colors. Plugging the ansatz \eqref{Eq:AnsatzMetric} in  \eqref{Eq:Einstein} we get the following system of coupled differential equations:
\begin{equation}\label{Eq:Background}
\begin{split}
\frac{\zeta''}{\zeta}-\frac{4}{9}(\Phi')^2=&\,0,\\
f''-\frac{3\zeta'}{\zeta}f'-(\zeta\,A_0')^2=&\,0,\\
3\zeta''-12\frac{\zeta'^2}{\zeta}+\frac{3f'\zeta'}{f}
+\frac{2\,V-\zeta^4(A_0')^2}{2\,f\,\zeta}=&\,0,
\end{split}
\end{equation}
where the primes stand for total derivatives with respect to the variable $z$, $d/dz$. As can be seen, these equations involve the warp factor, dilaton field and its potential, horizon function, and the gauge field. On the other hand, the Maxwell equation is obtained plugging \eqref{Eq:AnsatzMetric} in \eqref{Eq:Maxwell}
\begin{equation}
\left(\frac{A_0'}{\zeta}\right)'=0,
\end{equation}
which solution is given by $A_0'=c_1\,\zeta(z)$, where $c_1$ is the integration constant. The complete solution of the gauge field is given by
\begin{equation}\label{Eq:GaugeSol}
A_0(z)=c_2+c_1\,\int_{z}dx\,\zeta(x).
\end{equation}

\subsection{The holographic model}

In the following we introduce the holographic model we are going to work with here, which is the model originally proposed in Ref.~\cite{Gursoy:2007er}, Appendix G, and that was also used to investigate deconfined properties of $SU(N)$ Yang-Mills theories in Ref.~\cite{Caselle:2011mn}. In such a holographic model, the warp factor is given by 
\begin{equation}\label{Eq:WarpSol}
\zeta(z)=\frac{z}{\ell}e^{\Lambda^2z^2},
\end{equation}
where $\Lambda$ is a free parameter. Our motivation to use the analytic warp factor is the same as in Ref.~\cite{Caselle:2011mn}. Even though we do not know the explicit form of the dilaton potential, this warp factor allows us to investigate some general properties of the dual field theory in the strong coupling regime. It is worth mentioning that there are three ways to solve the coupled equations \eqref{Eq:Background}. The first one is to know the dilaton potential, one may get $\zeta$, $\Phi$, and $A_0$ solving the differential equations. The second one is to postulate the dilaton field, motivated by phenomenological constraints, one may get $\zeta$, $V$, and $A_0$ solving the corresponding differential equations. The third approach is to postulate the warp factor, motivated by phenomenological constraints, one may get $\Phi$, $V$, and $A_0$ solving the equations.

In order to solve the field equations, we start with the gauge field. By plugging Eq.~\eqref{Eq:WarpSol} into Eq.~\eqref{Eq:GaugeSol} implies in a relation that may be integrated to give
\begin{equation}\label{Eq:GaugeSol2}
A_0(z)=c_2+\frac{c_1}{\ell}\frac{\left(e^{\Lambda^2z^2}-1\right)}{2\Lambda^2}, 
\end{equation}
where $c_1$ and $c_2$ are integration constants. We may fix them by comparing the asymptotic form of \eqref{Eq:GaugeSol2}, close to the boundary $z\to 0$, with the corresponding Reissner-Nordstr\"om AdS (RNAdS) solution, which takes the form $A_0=\mu-Qz^2+\cdots$ \cite{Park:2009nb,Park:2011qq,Sin:2007ze}). Thus, we get $c_2=\mu$, $c_1=-2\ell\,Q$, and the gauge field is may be written as
\begin{equation}\label{Eq:GaugeSol3}
A_0(z)=\mu+\frac{Q}{\Lambda^2}\left(1-e^{\Lambda^2z^2}\right),
\end{equation}
where $\mu$ is interpreted as the chemical potential, and $Q$ is a parameter related to the electric charge of the black hole. The gauge field should be regular at the horizon, i.e., it may be imposed that $A_0(z_h)=0$. This condition allows us to find a relation between $Q$ and $\mu$, namely,
\begin{equation}\label{Eq:RelationQmu}
Q=\frac{\Lambda^2 \mu}{e^{\Lambda^2z_h^2}-1}.
\end{equation}
By substituting the last relation into \eqref{Eq:RelationQmu}, we get the final analytic solution for the gauge field
\begin{equation}
A_0(z)=\mu\frac{e^{\Lambda^2z^2}-e^{\Lambda^2z_h^2}}{1-e^{\Lambda^2zh^2}}.
\end{equation}
From the expansion of this result close to the boundary, $A_0=\mu-\rho z^2+\cdots$, and using the holographic dictionary we read of the quark density vacuum expectation value \footnote{In the literature $\rho$ is also known as baryon density.} in the dual field theory \cite{He:2013qq}
\begin{equation}\label{Eq:Rhomu}
\rho=-\frac{\Lambda^2 \mu}{1-e^{\Lambda^2 z_h^2}}.
\end{equation}
Thus, the relation between the parameter $Q$ and quark density is $\rho=Q$. The current-source coupling in the dual field theory is of the form $S=\int dx^4 \mu J^0$, where $\rho=\langle J^0\rangle$.

Proceeding with the process of solving the field equations, we plug Eqs.~\eqref{Eq:WarpSol} and \eqref{Eq:GaugeSol3} into the second equation of the system  \eqref{Eq:Background} to get a differential equation for the horizon function $f(z)$, whose solution is given by
\begin{equation}\label{Eq:HorizonFunct}
f=c_4+ c_3e^{3\Lambda^2z^2}\!\left(3\Lambda^2z^2-1\right)+\frac{Q^2e^{4\Lambda^2z^2}\!\left(4\Lambda^2z^2-1\right)}{16\,\ell^2\,\Lambda^6}.
\end{equation}
We fix the integration constants $c_3$ and $c_4$ with the conditions $f(0)=1$ and $f(z_h)=0$, what yields
\begin{equation}
\begin{split}
c_3=&\,-\frac{144\,\ell^2\Lambda^6+9Q^2\left[1+e^{4\Lambda^2z_h^2}(4\Lambda^2z_h^2-1)\right]}{144\,\ell^2\Lambda^6\left[1+e^{3\Lambda^2z_h^2}\left(3\Lambda^2z_h^2-1\right)\right]},\\
c_4=&\,\frac{e^{3\Lambda^2z_h^2}(16\,\ell^2\Lambda^6+Q^2)(3\Lambda^2z_h^2-1)}{16\,\ell^2\Lambda^6\left[1+e^{3\Lambda^2z_h^2}\left(3\Lambda^2z_h^2-1\right)\right]}\\
&+\frac{Q^2e^{4\Lambda^2z_h^2}(1-4\Lambda^2z_h^2)}{16\,\ell^2\Lambda^6\left[1+e^{3\Lambda^2z_h^2}\left(3\Lambda^2z_h^2-1\right)\right]}.
\end{split}
\end{equation}

The relation between the parameter $Q$ and the electric charge $q$ may be obtained by expanding Eq.~\eqref{Eq:HorizonFunct} close to the boundary where this expression reduces to the RNAdS solution. Thus, we get
\begin{equation}\label{Eq:RelationQq}
Q^2=3\ell^2\,q^2,
\end{equation}
where $q$ is the charge of the black hole. 

It is also worth mentioning that the limit of zero $\Lambda$ of Eq.~\eqref{Eq:HorizonFunct} is
\begin{equation}\label{Eq:RNSol}
f(z)=1-\left( {1}+q^2\,z_h^6\right)\frac{z^4}{z_h^4}+q^2\,z^6,
\end{equation}
which is the horizon function of the RNAdS black hole solution. On the other hand, by expanding Eq.~\eqref{Eq:HorizonFunct} close to the boundary, i.e., in the limit of small $z_h$ and $z$, we obtain
\begin{equation}\begin{split}\!\!
f=& 1+\left(-\frac{1}{z_h^4}+\frac{2\Lambda^2}{z_h^2}-q^2z_h^2+\mathcal{O}(z_h^3, \Lambda^4)\right)z^4\\
& +\left(q^2-\frac{2\Lambda^2}{z_h^4}+\frac{4\Lambda^4}{z_h^2}+\mathcal{O}(z_h^3,\Lambda^6)\right)z^6+\mathcal{O}(z^7), \end{split}
\end{equation}
where $\mathcal{O}(\Lambda, z_h)$ represent subleading contributions. Interestingly, the RNAdS solution is recovered in the limit of zero $\Lambda$. Then, we may see this equation as a deformed RNAdS black hole solution.

The dilaton field may be obtained by plugging Eq~\eqref{Eq:WarpSol} into the first equation of the system \eqref{Eq:Background} and by integrating the resulting relation. Then, the dilaton takes the form \cite{Gursoy:2007er}
\begin{equation}
\Phi=\frac{3z\Lambda}{4}\sqrt{6+4z^2\Lambda^2}+\frac{9}{4}\ln{\left[\frac{2z\Lambda+\sqrt{6+4z^2\Lambda^2}}{\sqrt{6}}\right]}.
\end{equation}
By expanding the function $\Phi(z)$ given by the last equation close to the boundary, we get
\begin{equation}\label{Eq:DilUV}
\Phi=\phi_1 z+\phi_3z^3+\cdots,
\end{equation}
where $\phi_1=3\Lambda/\sqrt{6}$ is interpreted as the source, while $\phi_3=\Lambda^3/\sqrt{6}$ is interpreted as the vacuum expectation value in the dual field theory. Considering the general expansion of the dilaton field close to the boundary $\Phi=\phi_1z^{4-\Delta}(1+\cdots)+\phi_3z^{\Delta}(1+\cdots)$, where $\Delta$ is the conformal dimension of the operator dual to the dilaton, while ellipses denote higher powers in $z$. Comparing this expression with \eqref{Eq:DilUV} we may conclude that $\Delta=3$, which means that $\Phi$ is dual to an operator of dimension three.

In turn, the dilaton potential may be calculated directly from the last equation of the system \eqref{Eq:WarpSol}. Taking analytical solutions of the background functions $f(z)$, $\zeta(z)$, and $A_0(z)$, and expanding close to the boundary it follows
\begin{equation}\label{Eq:DilPotUV}
V=\frac{12}{\ell^2}+\frac{54\Lambda^2 z^2}{\ell^2}+\cdots
\end{equation}
Writing this in terms of the dilaton field \eqref{Eq:DilUV}, the potential becomes
\begin{equation}
V=\frac{12}{\ell^2}-\frac{4}{3}M_{\Phi}^2 \Phi^2+\cdots,
\end{equation}
where $M_{\Phi}^2\ell^2=\Delta(\Delta-4)$ is the mass of the dilaton field. By setting $\Delta=3$ we recover \eqref{Eq:DilPotUV}, showing the consistency of the expansion \eqref{Eq:DilUV}.

\subsection{Thermodynamics}

The knowledge of the horizon function allows the calculation of the Hawking temperature, which is defined by
\begin{equation}
T=-\frac{1}{4\pi}\partial_{z}f(z)\bigg{|}_{z=z_h}.
\end{equation}
Since the present holographic model presents an analytic expression for the warp factor, cf. Eq.~\eqref{Eq:WarpSol}, the expression of the temperature may be expressed in exact form. It reads
\begin{widetext}
\begin{equation}\label{Eq:Temperature}
T=\frac{3 z_h^3\,e^{3\Lambda^2 z_h^2} \left[48\,\Lambda^6 -q^2\left(16\,e^{\Lambda^2 z_h^2}+e^{4\Lambda^2 z_h^2}\left(12\Lambda^2 z_h^2-7\right)-9\right)\right]}{32\pi\,\Lambda^2\big[1+e^{3\Lambda^2 z_h^2}\left(3\Lambda^2 z_h^2-1\right)\big]}.
\end{equation}
\end{widetext}
The dependence of the temperature on the chemical potential is obtained by plugging Eqs.~\eqref{Eq:RelationQq} and \eqref{Eq:RelationQmu} into \eqref{Eq:Temperature}. 

The next important thermodynamic variable to be easily calculated is the black hole entropy, which is determined by means of the Bekenstein-Hawking formula,
\begin{equation}
S=\frac{\mathcal{A}}{4 G_5}=\frac{2\pi V_3}{\kappa^2\,\zeta^3(z_h)}
\end{equation}
where $\mathcal{A}$ is the transverse area and $V_3$ is the three-dimensional space volume. Substituting Eq.~\eqref{Eq:WarpSol} into the last expression, it is straightforward to show that the entropy density $(s=S/V_3)$ may be written as
\begin{equation}\label{Eq:Entropy}
s=\frac{2\pi\ell^3}{\kappa^2\,z_h^3}e^{-3\Lambda^2 z_h^2}.
\end{equation}

So far we have calculated the temperature and entropy density as a function of the horizon radius and the electric charge (or chemical potential). In the way to obtain the EoS of the system, we note that the first law of thermodynamics in the presence of a chemical potential may be rewritten as \cite{DeWolfe:2010he}
\begin{equation}
d\mathcal{F}=-s\, dT-\rho\, d\mu,
\end{equation}
where $\mathcal{F}$ is the free energy density and $\rho$ is the quark density. The free energy may be calculated through the relation
\begin{equation}\label{Eq:FreeEnergy}
\mathcal{F}=\int_{z_h}^{\infty} s(\tilde{z}_h)\left(\frac{dT(\tilde{z}_h)}{d \tilde{z}_h}\right)d \tilde{z}_h.
\end{equation}
In the last expression, we implicitly imposed the condition that the free energy in the limit of arbitrarily small black hole, i.e., in the limit $z_h\to \infty$, the free energy of the dual thermal gas is zero \cite{He:2013qq}. 

Once we have got the free energy, the pressure is also known. It is given by
\begin{equation}\label{Eq:Pressure}
p=-\mathcal{F}.
\end{equation}

Finally, the energy density is calculated through the thermodynamic relation $\epsilon=\mathcal{F}+Ts+\mu\rho$. That completes the task of building an EoS that, through the holographic dictionary, will describe a dual fluid to the charged-dilaton AdS black hole.

\subsection{Asymptotic analysis}

Let us calculate the asymptotic behavior of the thermodynamic variables. In this section we work on the branch where there are no instabilities, i.e., close to the boundary, see Fig.~\ref{Fig:Tzh}. Then, expanding Eq.~\eqref{Eq:Temperature} close to the boundary (which means that we are considering the big black holes branch) and plugging the relation \eqref{Eq:RelationQq} the temperature becomes
\begin{equation}
T=\frac{1}{\pi z_h}+\frac{\Lambda^2z_h}{\pi}+\frac{\Lambda^4 z_h^3}{4\pi}-\left(\frac{q^2}{2\pi}+\frac{\Lambda^6}{20\pi}\right)z_h^5+\mathcal{O}(z_h^7).
\end{equation}
It is worth mentioning that in the limit of vanishing $\Lambda$ we recover the RNAdS expression for the temperature. One more interesting relationship is the one relating the temperature and chemical potential, which is obtained plugging \eqref{Eq:RelationQmu} and \eqref{Eq:RelationQq} in \eqref{Eq:Temperature}, then, expanding close to the boundary 
\begin{equation}\label{Eq:Temperature2}
\begin{split}
T=&\,\frac{1}{\pi z_h}+\left(\frac{\Lambda^2}{\pi}-\frac{\mu^2}{6\ell^2}\right)z_h+\left(\frac{\Lambda^4}{4\pi}-\frac{3\Lambda^2\mu^2}{8\pi \ell^2}\right)z_h^3\\
&-\left(\frac{\Lambda^6}{20\pi}+\frac{151\Lambda^4\mu^2}{360\pi\ell^2}\right)z_h^5+\mathcal{O}(z_h^7).
\end{split}
\end{equation}

The RNAdS solution is obtained by setting $\Lambda$ to zero and replacing $\mu/z_h^2=Q$. Now, we may invert the asymptotic series \eqref{Eq:Temperature2} such that the expression for $z_h$ becomes
\begin{equation}\label{Eq:zhT}
z_h=\frac{1}{\pi T}+\left(\frac{\Lambda^2}{\pi^3}-\frac{\mu^2}{6\pi^3\ell^2}\right)\frac{1}{T^3}+\mathcal{O}(1/T^5).
\end{equation}
This result will be useful when we calculate the thermodynamic variables as a function of the chemical potential and temperature.

Let us now calculate the asymptotic expressions for the quark density, entropy, and free energy density close to the boundary, i.e., in the high-temperature regime. Then, plugging \eqref{Eq:zhT} in \eqref{Eq:Rhomu} we get 
\begin{equation}
\rho=\frac{\mu^3}{3\ell^2}-\frac{5\Lambda^2 \mu}{2}+\pi^2 T^2 \mu+\mathcal{O}(\Lambda^4 \mu/T^2).
\end{equation}
Whereas the entropy density, Eq.~\eqref{Eq:Entropy}, becomes
\begin{equation}
\begin{split}
s&=\frac{2\pi^4\ell^3 T^3}{\kappa^2}+\frac{\pi^2\ell(\mu^2-12\Lambda^2\ell^2)T}{\kappa^2}\\
&+\mathcal{O}\left(\Lambda^4/T,{\Lambda^2\mu^2/T},{\mu^4/T}\right).
\end{split}
\end{equation}
In turn, the free energy density \eqref{Eq:FreeEnergy} is given by
\begin{equation}\label{Eq:FreeEnerHighT}
\begin{split}
\mathcal{F}=&-\frac{\pi^4\ell^3 T^4}{2\kappa^2}+\frac{\pi^2\ell}{6\kappa^2}\left(36\Lambda^2\ell^2-7\mu^2\right)T^2\\
&-\frac{9\ell^3\Lambda^4}{\kappa^2}+\frac{10\ell\Lambda^2\mu^2}{3\kappa^2}-\frac{11\mu^4}{36\kappa^2\ell}+\cdots,
\end{split}
\end{equation}
while the pressure is the negative of the free energy. It is worth pointing out that the free energy \eqref{Eq:FreeEnerHighT} splits out in a piece depending on the temperature, another depending on the chemical potential and the last one depending on a mixture. Following an argument similar to the one of Ref.~\cite{Hoyos:2016zke} (see also \cite{Annala:2017tqz}), neglecting the piece depending on the temperature, then, we may calculate an EoS, which does not depend on the temperature. Thus, the free energy we are going to work with is given by
\begin{equation}
\mathcal{F}=-\frac{11\mu^4}{36\kappa^2\ell}+\frac{10\ell\Lambda^2\mu^2}{3\kappa^2}-\frac{9\ell^3\Lambda^4}{\kappa^2}.
\end{equation}
It is worth mentioning that we do not impose any condition on $\mu$ and $\Lambda$ to get this result. The free energy has the same mathematical structure as one expression proposed in Ref.~\cite{Alford:2004pf}, where the authors used a phenomenological construction to investigate the quark matter stars. Hence, the results we obtain support the heavy quarks interpretation we are considering in this work. In turn, the pressure is given by
\begin{equation}\label{Eq:pHighT}
p=\frac{11\mu^4}{36\kappa^2\ell}-\frac{10\ell\Lambda^2\mu^2}{3\kappa^2}+\frac{9\ell^3\Lambda^4}{\kappa^2}.
\end{equation}
Then, at large densities the energy density is obtained using $\epsilon=\mu\partial_{\mu}p-p$, 
\begin{equation}\label{Eq:EHighT}
\epsilon=\frac{11\mu^4}{12\kappa^2\ell}-\frac{10\ell\Lambda^2\mu^2}{3\kappa^2}-\frac{9\ell^3\Lambda^4}{\kappa^2},
\end{equation}
rewriting the last expression as a function of the pressure
\begin{equation}\label{Eq:EoSHighT}
\epsilon=3p+\frac{40\ell^{3/2}\Lambda^2\sqrt{11\kappa^2 p+\ell^3\Lambda^4}}{11\kappa^2}+\frac{4\ell^3\Lambda^4}{11\kappa^2}.
\end{equation}
So far, we do not impose any restrictions on the chemical potential, i.e., if it is large or small. Expanding \eqref{Eq:EoSHighT} in the regime of high pressures.
\begin{equation}\label{Eq:EoSHighP}
\epsilon=3p+\frac{40\ell^{3/2}\Lambda^2}{\sqrt{11}\kappa}\sqrt{p}+\frac{4\ell^3\Lambda^4}{11\kappa^2}+\mathcal{O}(1/\sqrt{p}).
\end{equation}
It is worth mentioning that Eq.~\eqref{Eq:EoSHighP} has the same mathematical form, unless the constant, that the EoS obtained in the holographic model of Ref.~\cite{Hoyos:2016zke}.

\subsection{Extremal solution}
In the previous section we have calculated an analytic EoS by approximating the thermodynamics variables close to the boundary, i.e., in the high-temperature regime. However, from a pragmatic point of view, it is more convenient to find out an EoS in the zero temperature case. A way for reaching this requirement in the holographic model is to consider the extremal solution. This solution is obtained when the Hawking temperature \eqref{Eq:Temperature} vanishes. Hence, by solving the equation $T=0$ we may express the parameter $q$ as a function of  $\Lambda$ and $z_h$ in the form 
\begin{equation}\label{Eq:qExtremal}
q=\frac{4\sqrt{3}\Lambda^3}{\left(-9+16e^{\Lambda^2 z_h^2}+e^{4\Lambda^2 z_h^2}(12\Lambda^2 z_h^2-7)\right)^{1/2}}.
\end{equation}
The relation between the chemical potential $\mu$, $\Lambda$, and $z_h$ in the extremal case is obtained by plugging relation \eqref{Eq:qExtremal} in \eqref{Eq:RelationQq},
\begin{equation}
\mu=\frac{12\,\ell\,\Lambda\left(e^{\Lambda^2 z^2_h}-1\right)}{\left(-9+16e^{\Lambda^2 z_h^2}+e^{4\Lambda^2 z_h^2}(12\Lambda^2 z_h^2-7)\right)^{1/2}},
\end{equation}
where Eq.~\eqref{Eq:RelationQmu} was also employed. It is convenient to get a relation of $z_h$ as a function of the chemical potential, for that reason we may expand the last equation close to the boundary, i.e., $z_h=0$,
\begin{equation}\label{Eq:muzh}
\mu=\frac{\sqrt{6}\,\ell}{z_h}-\frac{5\sqrt{6}\,\ell\Lambda^2 z_h}{8}+\frac{149\,\ell \Lambda^4 z_h^3}{320\sqrt{6}}+\mathcal{O}(z_h^5)
\end{equation}
Inverting the last asymptotic series we get
\begin{equation}\label{Eq:zhmu}
z_h=\frac{\sqrt{6}\,\ell}{\mu}-\frac{15\sqrt{6}\,\ell^3\Lambda^2}{4\mu^3}+\mathcal{O}(1/\mu^5).
\end{equation}
Then, plugging \eqref{Eq:zhmu} in \eqref{Eq:Entropy} we get the entropy. However, as one of the aims of this paper is to get an analytic EoS we expand the entropy in the region of large densities, this approximation gives us the result
\begin{equation}\label{Eq:ExtremalEntropy}
s=\frac{\pi \mu^3}{3\sqrt{6}\kappa^2}-\frac{3\sqrt{6}\pi\ell^2\Lambda^2\mu}{8\kappa^2}+\frac{159\sqrt{6}\pi\ell^4\Lambda^4}{16\kappa^2 \mu}+\mathcal{O}(1/\mu^3)
\end{equation}
In the regime of large densities terms depending on $\mathcal{O}(1/\mu)$ may be neglected. We observe that Eq. \eqref{Eq:ExtremalEntropy} in the $\Lambda$ zero limit reduces to the RNAdS extremal result, cf. \eqref{Eq:ExtremalAdSRNSol} in Appendix \ref{AppendixA}. The subleading terms in Eq. \eqref{Eq:ExtremalEntropy} may be interpreted as deformations of the RNAdS extremal solution. Implementing the same procedure for the free energy we get
\begin{equation}
\mathcal{F}=-\frac{\mu^4}{12\ell\,\kappa^2}-\frac{75\ell^3\Lambda^4}{32\kappa^2}+\mathcal{O}(\log (\sqrt{6}\ell/\mu),1/\mu^2).
\end{equation}
The pressure is given by $p=-\mathcal{F}$. It is worth mentioning that in the $\Lambda$ zero limit we recover the RNAdS extremal solution, see Eq. \eqref{Eq:ExtremalAdSRNSol} in Appendix \ref{AppendixA}. In turn, the EoS is given by
\begin{equation}
\epsilon=3p-\frac{75\ell^3 \Lambda^4}{8\kappa^2}.
\end{equation}
The subleading term may be interpreted as a deformation of the conformal theory, which EoS is given by $\epsilon=3p$.

To finish this section we are going to rewrite the metric \eqref{Eq:AnsatzMetric} close to the horizon. Then, expanding the horizon function \eqref{Eq:HorizonFunct} becomes
\begin{equation}
f\approx \mathcal{K}\left(12\frac{\tilde{z}^2}{z_h^2}+\left(28+56z_h^2\Lambda^2\right)\frac{\tilde{z}^3}{z_h^3}+\mathcal{O}(\tilde{z}^4/z_h^4)\right),
\end{equation}
where $\tilde{z}=z-z_h$, the constant $\mathcal{K}$ is given by
\begin{equation}
\mathcal{K}=\frac{24\Lambda^6z_h^6\,e^{4z_h^2\Lambda^2}}{-9+16e^{z_h^2\Lambda^2}+e^{4z_h^2\Lambda^2}\left(12z_h^2\Lambda^2-7\right)}.
\end{equation}
Then, close to the horizon the metric becomes
\begin{equation}
\begin{split}
ds^2=&-12\frac{\mathcal{K}e^{-z_h^2\Lambda^2}\ell^2\tilde{z}^2}{z_h^4}dt^2+\frac{\ell^2e^{-z_h^2\Lambda^2}}{12\mathcal{K}\tilde{z}^2}d\tilde{z}^2\\
&+\frac{\ell^2e^{-z_h^2\Lambda^2}}{z_h^2}dx_idx^i.
\end{split}
\end{equation}
Finally, after defining the new variable $\tilde{z}=\hat{r}z_h^2/\ell^2$ the last equation reads
\begin{equation}\label{Eq:ExtremalMetric}
\begin{split}
ds^2=&-12\frac{\mathcal{K}e^{-z_h^2\Lambda^2}\hat{r}^2}{\ell^2}dt^2+\frac{\ell^2e^{-z_h^2\Lambda^2}}{12\mathcal{K}\hat{r}^2}d\hat{r}^2\\
&+\frac{\ell^2e^{-z_h^2\Lambda^2}}{z_h^2}dx_idx^i.
\end{split}\end{equation}
The point is that in the limit of zero $\Lambda$, $\mathcal{K}=1$ and the metric \eqref{Eq:ExtremalMetric} reduces to the $AdS_2\times \mathbb{R}^3$ (see Appendix \ref{AppendixA}). The emergence of this geometry is related to the emergence of quantum criticality in the dual field theory \cite{Faulkner:2009wj}. Hence, we may interpret the metric \eqref{Eq:ExtremalMetric} as a deformation of the $AdS_2\times\mathbb{R}^3$ geometry due to the presence of the parameter $\Lambda$.

\section{Thermodynamics: Numerical results}
\label{Sec:NumericalResults}

We start this section by fixing the parameters of the model.
In order to do that we first search for a critical value of the chemical potential, $\mu_c$, for which the temperature is a monotonic decreasing function of the horizon radius ${z_h}$. This value may be calculated by solving the equations $\partial_{z_h}T(z_h,\mu,\Lambda)=0$ and $\partial^2_{z_h}T(z_h,\mu,\Lambda)=0$, which are the conditions for the curve having no maximum neither minimum points. These two equations together with Eq.~\eqref{Eq:Temperature} form the system to be solved. 
Since the model presents five unknown parameters, namely, $ \kappa$, $q^2$ (or $\mu$), $\Lambda$, $z_h$, $T$, additional conditions are need. We set the parameter $\kappa$, as usual, by $\kappa=2\pi/N$, with $N=3$.Additionally, here we set $T=T_c=0.270$ GeV, where $T_c$ plays the role of the critical temperature. Then we are left with a closed system to be solved, three equations for the three unknown quantities, namely, $q^2$ (or $\mu$), $\Lambda$, and $z_h$. In fact, with such choices it is possible to solve numerically the system of equations mentioned above to get $\mu\equiv \mu_c=0.095$ GeV, $\Lambda \equiv\Lambda_c=0.402$ GeV, and $z_{h_c}=2.434$. It is worth pointing out that the value of $\Lambda$ fixed in this way is of the same order of magnitude as the value found by means of the vector meson spectra in the holographic soft wall model \cite{Gherghetta:2009ac} (see also \cite{Ballon-Bayona:2020qpq}).

\begin{figure}[ht]
\centering
\includegraphics[width=8cm]{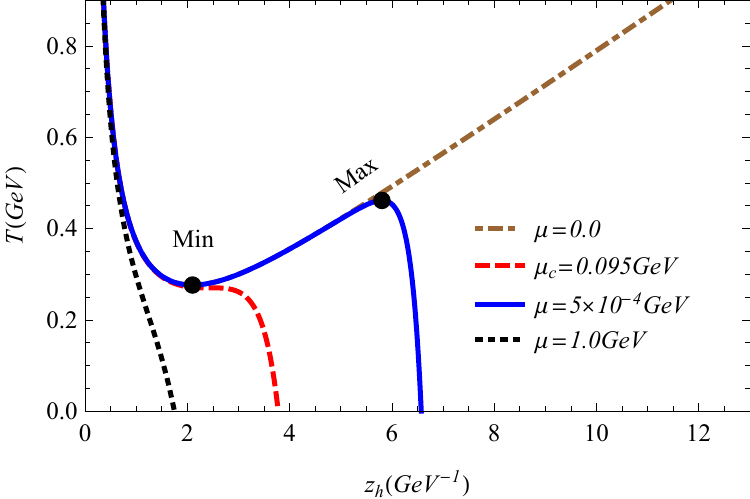}
\caption{The figure shows the temperature in terms of the horizon radius for different values of the chemical potential. The other parameters of the model are fixed as $\kappa=2\pi/N$, with $N=3$, and $\Lambda=0.402\,\text{GeV}$.}
\label{Fig:Tzh}
\end{figure}

In the remaining of this section, the parameter $\Lambda$ is fixed to its critical value corresponding to the critical temperature $T_c=0.270$ GeV, i.e., $\Lambda=0.402$ GeV, and the parameter $\kappa=2\pi/3$ also is kept fixed. With the two parameters being fixed, and taking different values of the chemical potential, we may analyze the behavior of the temperature, cf. Eq.~\eqref{Eq:Temperature}, as a function of the horizon radius $z_h$. Some plots of the resulting functions are displayed in Fig.~\ref{Fig:Tzh}, where it is seen also the strong dependence of $T=T(z_h)$ on the value of the chemical potential. In particular, the figure displays the behavior of the temperature for the critical value $\mu_c$, see the red line in Fig.~\ref{Fig:Tzh}, and for three other selected values of the chemical potential, as indicated. 

Still concerning Fig.~\ref{Fig:Tzh}, it is verified that for any arbitrary value of the chemical potential $\mu$ in the interval $0< \mu <\mu_c$, the graph for the temperature presents a local minimum at $z_{h_{\text{Min}}}$, say, and a local maximum at $z_{h_{\text{Max}}}$, see the case for $\mu=5\times10^{-4} \,{\rm GeV} $ (brown line) shown in the figure. A system (a black hole) with configuration such that $z_h<z_{h_{\text{Min}}}$, the big black hole branch, is stable from the thermodynamic point of view. In turn, the branch belonging to the region $z_{h_{\text{Min}}}\leq z_h\leq z_{h_{\text{Max}}}$ represents an unstable phase where the specific heat becomes negative, as we shall see below. Finally, a black hole whose configuration is in the region $z_h> {z}_{h_{\text{Max}}}$ is also thermodynamically stable.  For additional discussions on this subject see, for instance, Ref.~\cite{He:2013qq} (see also the recent paper \cite{Ballon-Bayona:2020xls}).

\begin{figure}[ht]
\centering
\includegraphics[width=8cm]{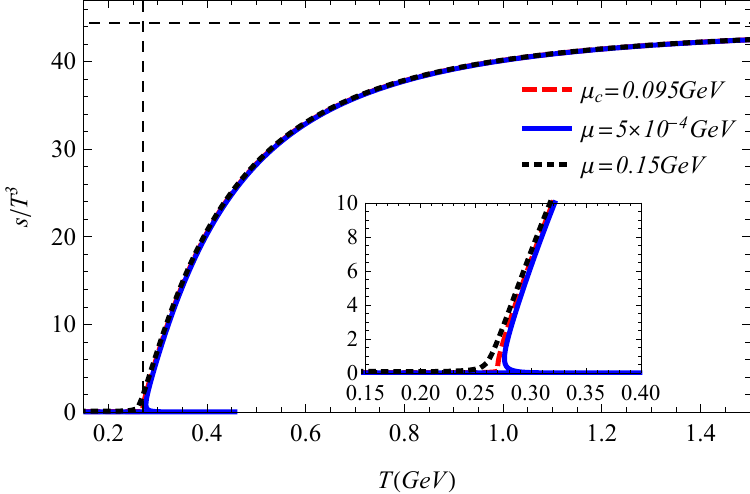}
\caption{The figure shows $s/T^3$ as a function of the temperature for different values of the chemical potential. We observe that the values asymptotically approach the conformal value, the horizontal black dashed line, for which $s/T^3=\pi^2\ell^2N^2/2$. We have set $N=3$ and $\Lambda=0.402\,\text{GeV}$. The vertical dashed line is drawn for $T=T_c=0.270 \,\text{GeV}$.}
\label{Fig:ST}
\end{figure}

Figure~\ref{Fig:ST} shows a plot of $s/T^3$ as a function of the temperature for three different values of the chemical potential, $\mu=\mu_c=0.095$ Gev (dashed red line), $\mu= 5\times 10^{-4}$ GeV (solid blue line), and $\mu =0.15$ GeV (dotted black line). The asymptotic behaviour of the function $s/T^3$ is the same the three cases, approaching  the conformal limit $s/T^3 = \pi^2\ell^2N^2/2$ at high temperatures. As seen in the inset of that figure, the main difference is found close to the critical temperature $T=T_c=0.270$ GeV.The function $s/T^3$ stars to grow just above $T_c$, at $T=T_c$, and just below $T_c$, respectively, for $\mu=\mu_c = 0.095$ Gev, $\mu= 5\times 10^{-4}$ GeV, and $\mu =0.15$ GeV.

\begin{figure}[ht!]
\centering
\includegraphics[width=8cm]{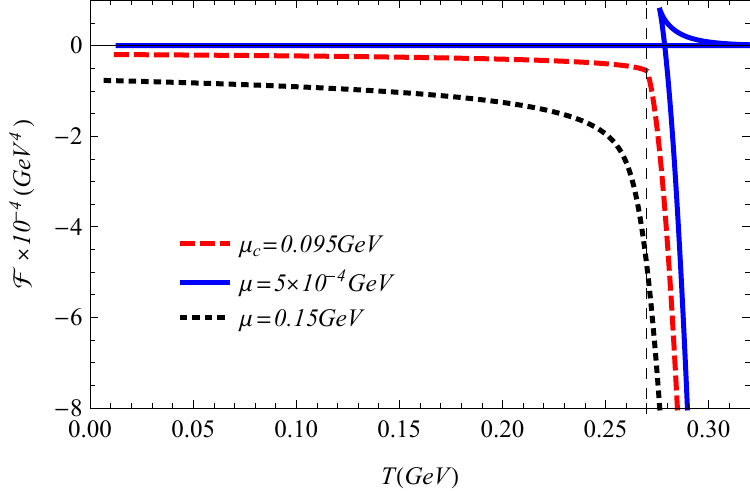}
\vskip .2cm 
\includegraphics[width=8cm]{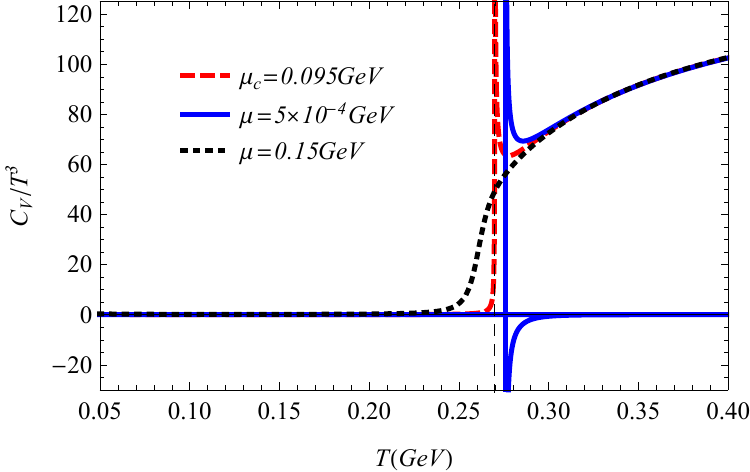}
\caption{Top: The free energy $\cal F$ as a function of the temperature for different values of the chemical potential: $\mu<\mu_c$ (solid blue line),  $\mu=\mu_c$ (dashed red line) and $\mu_c<\mu$ (dotted black line).\\
Bottom: The specific heat $C_V/T^3$ as a function of the temperature for different values of the chemical potential: $\mu<\mu_c$ (solid blue line),  $\mu=\mu_c$ (dashed red line) and $\mu_c<\mu$ (dotted black line). In both figures, the vertical line represents $T_c=0.270\, \text{GeV}$.}
\label{Fig:FTCT}
\end{figure}

Another important thermodynamic function, the free energy density $\cal F$ (or grand potential in the grand canonical ensemble) is displayed in the top panel of Fig.~\ref{Fig:FTCT}.  As can be seen, for $0<\mu<\mu_c$ (see the solid blue line), $\cal F$ is a multivalued function of the temperature. This behavior is a characteristic feature of a first-order phase transition, and it may be interpreted as representing a transition between the big black hole to the small black hole phases.  In turn, in the case, $\mu=\mu_c$ (dashed red line), the free energy is a smooth decreasing function of the temperature, except at the point $(T_c,\mathcal{F}_c)$, where the derivative is not defined. This is the critical point. Meanwhile, for $\mu>\mu_c$ (dotted black line), the free energy is a smooth decreasing function of the temperature. In the next section, we present the phase diagram and investigate a few more details on this subject. 

In turn, the specific heat is defined as
\begin{equation}
C_V=T\left(\frac{\partial s}{\partial T}\right)_{\mu}.
\end{equation}
The numerical results for the specific heat as a function of the temperature are displayed in the bottom panel of Fig.~\ref{Fig:FTCT} for different values of the chemical potential. We can see that, when $\mu<\mu_c$ the specific heat has a negative branch and is multivalued, representing the unstable phase arising between $z_{\text{Min}}<z_h<z_{\text{Max}}$ in Fig.~\ref{Fig:Tzh}. In turn, for $\mu\geq\mu_c$ the specific heat is always positive, representing the stable black hole phase.

\begin{figure}[ht]
\centering
\includegraphics[width=8cm]{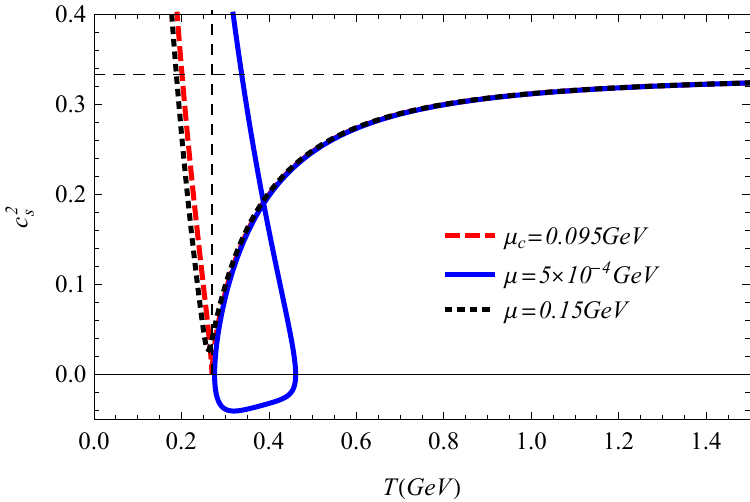}
\vskip .2cm
\includegraphics[width=8cm]{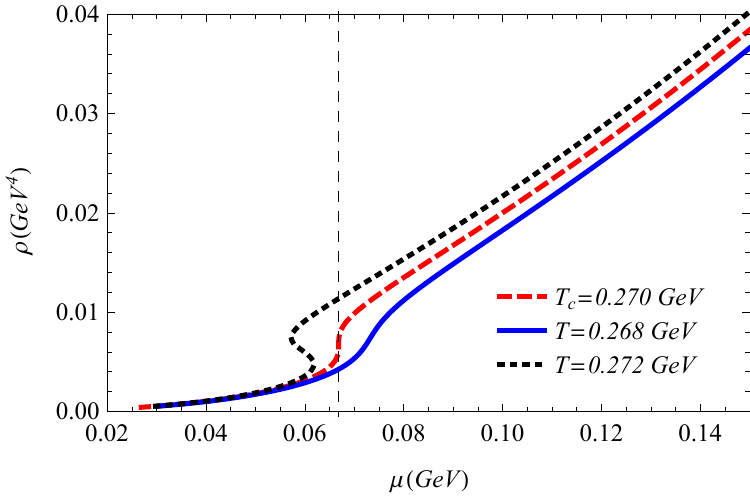}
\caption{Top: The figure shows the square of the speed of sound as a function of the temperature for different values of the chemical potential: $\mu<\mu_c$ (solid blue line),  $\mu=\mu_c$ (dashed red line) and  $\mu_c<\mu$ (dotted black line). Vertical dashed line represents $T_c=0.270\, \text{GeV}$, while the horizontal dashed line $c_s^2=1/3$.\\ 
Bottom: The figure shows the quark density as a function of the chemical potential for different values of the temperature: $T<T_c$ (blue line),  $T=T_c$ (red line), and  $T_c<T$ (black line). Vertical dashed line represents $T_c=0.270\, \text{GeV}$.}
\label{Fig:CsTRu}
\end{figure}

The speed of sound may be calculated from the relation
\begin{equation}
c_s^2=\frac{\partial \ln T}{\partial \ln s}.
\end{equation}
The top panel of Fig.~\ref{Fig:CsTRu} shows the dependence of $c_s^2$ on the temperature for three different values of the chemical potential. Notice that, for $\mu<\mu_c$ (solid blue line), $c_s^2$ becomes negative in a given interval of temperature, that is the region where thermodynamic instabilities arise. This is related to the fact that the specific heat is multivalued and may have negative values. In fact, one has the relation
\begin{equation}
c_s^2=\frac{s}{C_V}.
\end{equation}
For $\mu=\mu_c$ (dashed red line), the square of the speed of sound touches the horizontal axis ($c_s^2=0$) at $T=T_c$. This fact is also related to the specific heat, which blows up at the critical temperature. Moreover, for $T<T_c$ (dotted black line), $c_s^2$ increases rapidly. In turn, for $\mu>\mu_c$, it has a nonzero minimum at $T<T_c$. Finally, it is worth mentioning that in the high-temperature regime $c_s^2$ approaches asymptotically to the conformal value, $c_s^2=1/3$, in all the cases investigated here.

In the same way, we may calculate the susceptibility, which is defined by
\begin{equation}
\chi=\left(\frac{\partial \rho}{\partial \mu}\right)_{T}.
\end{equation}
The numerical results concerning this thermodynamic function are displayed in the bottom panel of Fig.~\ref{Fig:CsTRu} where we plot the quark density as a function of the chemical potential, $\rho = \rho(\mu)$,  for three different values of the temperature. It is seen that, for $T<T_c$ (solid blue line), $\rho(\mu)$ is a smooth monotonically increasing function of the chemical potential. In turn, for $T=T_c$ (dashed red line) the quark density is as increasing function of the chemical potential, but the slope of the curve for $\rho(\mu$) blows up at $\mu\simeq 0.067$ GeV (see the vertical dashed black in the figure), which means that the susceptibility function has a singularity at $T=T_c$. Meanwhile, for $T>T_c$  (dotted black line), $ \rho(\mu)$ is a multivalued function and $\chi$ presents two singularities in the region $\mu\neq\mu_c$, i.e., the quark density is such that $\partial_{\mu}\rho$ blows up for two specific values of the chemical potential. 

\begin{figure}[ht!]
\centering
\includegraphics[width=8cm]{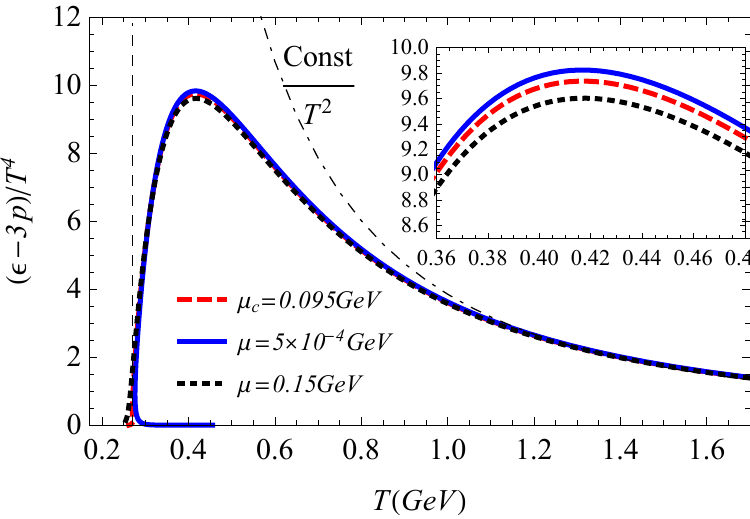}
\caption{The trace anomaly (or interaction measure) as a function of the temperature for different values of the chemical potential: for $\mu<\mu_c$ (blue line),  $\mu=\mu_c$ (red line), and for $\mu_c<\mu$ (black line). The vertical line is at $T_c=0.270\,\text{GeV}$, while the doted dashed line represents the function $\text{Const}/T^2$ where $\text{Const}\approx3.9$.}
\label{Fig:Trace}
\end{figure}

Finally, as we have all the thermodynamic variables on hands we may calculate the trace anomaly, $\epsilon-3p$. Our numerical results of the trace anomaly are displayed in Fig.~\ref{Fig:Trace}, where we plot it as a function of the temperature, as can be seen, there is a peak near to $T_c$, then, it decreases monotonically with the temperature. Whereas the maximum depends on the value of the chemical potential. As pointed out in Ref.~\cite{Caselle:2011mn}, the behavior of the trace anomaly goes like $\sim 1/T^2$ for $T>T_c$, we also notice this behavior in the region of $T>T_c$, for small values of the chemical potential.


\section{Phase diagram}
\label{Sec:PhaseDiagram}

\begin{figure}[ht!]
\centering
\includegraphics[width=8cm]{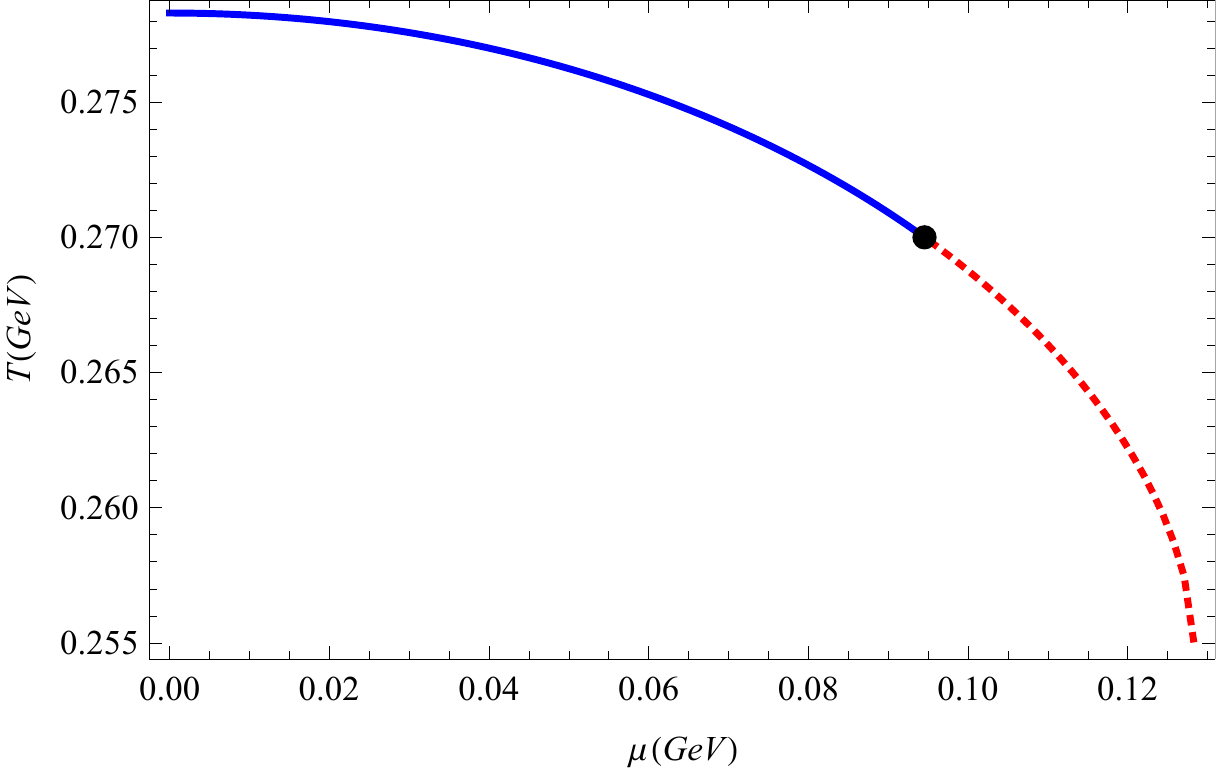}
\caption{
The phase diagram of the holographic model. The solid line represents the first-order phase transition region, the black dot point is the critical point, and the dashed line represents the crossover region.}
\label{Fig:Phase}
\end{figure}

Having described the thermodynamics and, after choosing an appropriate value for $\Lambda$, identified the critical point $(\mu_c, T_c)$, now we may draw the phase diagram. This is obtained by plotting the temperature as a function of the chemical potential. The phase diagram brings us information about the phase structure of the holographic model and shows the region in the parameter space where there are first-order, as well as crossover transitions in the model we are working with. This diagram also shows the location of the critical point, which lies at the end of the first-order transition line. The results of the numerical analysis are displayed in Fig.~\ref{Fig:Phase}, where we plot the temperature versus the chemical potential for $\Lambda=0.402\,\text{GeV}$ and $\kappa= 2\pi/3$. The solid line starts at $\mu=0$ and finishes at the critical point $(\mu_c, T_c)$, represented by the dark point in the figure. This line represents the region of first-order phase transitions. Above the critical point, the transition becomes crossover, which is represented in the figure by the dashed line. These results are in agreement with a holographic model describing the heavy quarks system, as discussed in Ref.~\cite{He:2013qq}.

So far we have studied the thermodynamic functions by fixing the parameter $\Lambda$ to its critical value, $\Lambda=0.402$ GeV, that corresponds to $T_c(z_h,\mu,\Lambda)= 0.270$ GeV and $\mu_c=0.095$ GeV. To complete the investigation it is worth verifying also the dependence of the critical temperature and of the critical potential on $\Lambda$. For that we may solve the equations $\partial_{z_h}T(z_h,\mu,\Lambda)=0$ and $\partial^2_{z_h}T(z_h,\mu,\Lambda)=0$ by choosing different values of $\Lambda$ time at a time. Thus, solving for $z_h$ the two equations and comparing the results we find a linear relation between the critical value of the chemical potential and $\Lambda$, i.e., $\mu_c=a_1 \Lambda+b_1$ with $a_1$ and $b_1$ being adjusting parameters. The relation between the critical temperature and $\Lambda$ is also approximately linear, $T_c=a_2\Lambda+b_2$, with $a_2$ and $b_2$ being adjusting parameters.

\section{Holographic compact stars}
\label{Sec:TOV}

\subsection{The heavy quarks zero temperature equation of state}

Here we analyze further the properties of the EoS for the matter at the core of hybrid stars, where the energy density is high enough to justify the use of a matter model based on the holographic approach. In this section, we relax the condition imposed on $\Lambda$ so that this parameter may be viewed as a free parameter of the model. This will allow us to investigate the dependence of the equation of state on $\Lambda$, and also to compare the results against other models available in the literature, especially those of Refs.~\cite{Hoyos:2016zke, Annala:2017tqz, Fadafa:2019euu}. The EoS we are working with here is given by Eq.~\eqref{Eq:EoSHighT}, which is obtained from Eqs.~\eqref{Eq:pHighT} and \eqref{Eq:EHighT}. In the sequence, we follow the procedure implemented in Ref.~\cite{Hoyos:2016zke}, which was also followed in Ref.~\cite{Fadafa:2019euu}. On the other hand, we fix the parameter $\Lambda$ to get zero pressure at $\mu=308.55$ MeV. In the present model, the zero pressure condition provides a relation between $\Lambda$ and $\mu$. In fact, from Eq.~\eqref{Eq:pHighT} we get
\begin{equation}
\Lambda=\frac{\mu}{\ell\sqrt{6}},
\end{equation}
from what follows $\Lambda=125.965$ MeV. 

\begin{figure}[ht!]
\centering
\includegraphics[width=8cm]{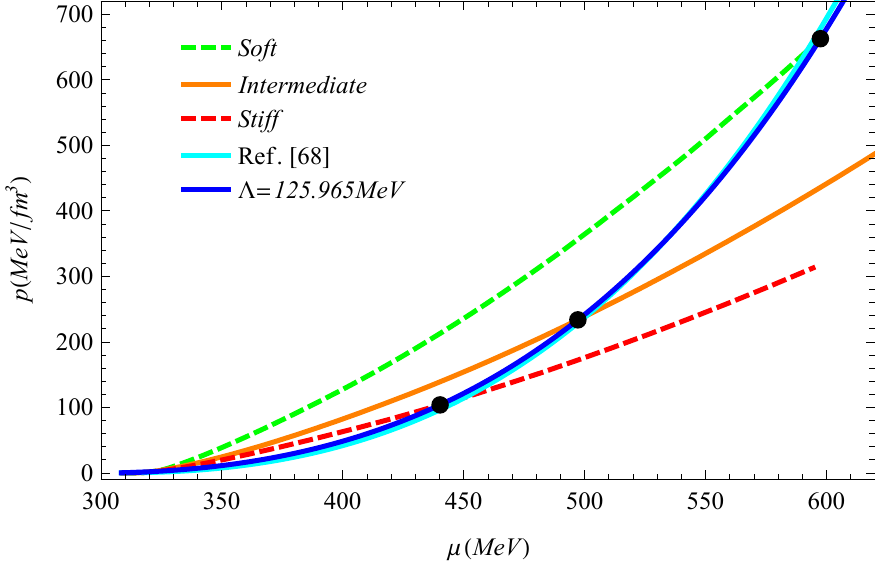}
\caption{The pressure as a function of the chemical potential for the holographic model (blue solid line) in comparison to the result of Ref.~\cite{Hoyos:2016zke} (cyan solid line). In drawing this figure we have chosen $\Lambda=125.965$ MeV (blue). The other lines represent the soft (dashed green), intermediate (solid orange), and stiff (dashed red) nuclear data of Ref.~\cite{Hebeler:2013nza}.}
\label{Fig:PMu}
\end{figure}

A plot of the pressure as a function of the chemical potential is displayed in Fig.~\ref{Fig:PMu} in comparison to the results of the holographic model of Ref.~\cite{Hoyos:2016zke}, and also to the nuclear matter EoSs of Ref.~\cite{Hebeler:2013nza}. In this figure, the heavy quark holographic EoS of the present work is plotted with a solid blue line, the nuclear matter EoSs results are plotted respectively with a dashed green line (soft nuclear EoS), a solid orange line (intermediate nuclear EoS), and by a dashed red line (stiff nuclear EoS), while the holographic EoS of Ref.~\cite{Hoyos:2016zke} is represented by a solid cyan line. Note that the solid blue and solid cyan lines are very close to each other. A matching procedure is implemented at the intersection points between the curve for the present holographic matter EoS, cf. Eq.~\eqref{Eq:EoSHighT}, and the curves for the nuclear matter EoSs. In this figure, the intersections are represented by black points, which are, respectively: $(\mu,\,p) = (440.272\,\text{MeV}$, $103.679\,\text{MeV}/\text{fm}^3)$ for the stiff case, $(\mu,\, p)= (497.252\,\text{MeV}, $ $233.466\,\text{MeV}/\text{fm}^3)$ for the intermediate case, and $(\mu,\, p) = (597.255\,\text{MeV},$ $662.592\,\text{MeV}/\text{fm}^3)$ for the soft EoS case.

As seen in Fig.~\ref{Fig:PMu}, the stiff matter EoS is dominant in the low densities region, then, a phase transition to the quark matter occurs at $\mu=440.272\,\text{MeV}$ (at the intersection with the solid blue line). This means that the star has an outer structure dominated by the stiff nuclear matter, while its central region is dominated by the quark matter EoS. As pointed out in Ref.~\cite{Hoyos:2016zke}, it is expected a first-order phase transition from nuclear to quark matter. The same is valid for the transition from intermediate and soft nuclear matter EoSs to quark matter EoS. Compared to the top-down holographic model of Ref.~\cite{Hoyos:2016zke} (solid cyan line), the transitions occur at lower densities in the cases of stiff and intermediate EoSs. On the other hand, in the case of the soft EoS, the transition occurs at higher densities when compared to the results of Ref.~\cite{Hoyos:2016zke}.

\begin{figure}[ht]
\centering
\includegraphics[width=8cm]{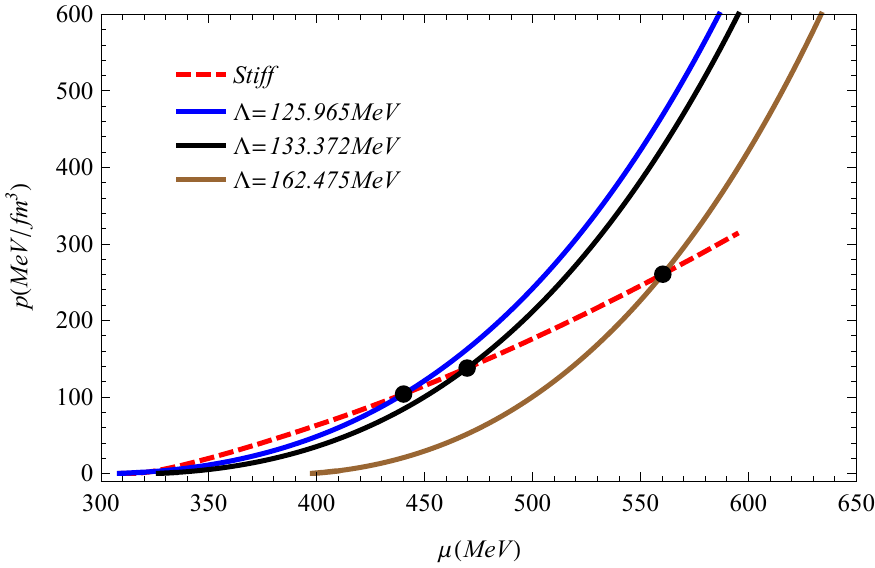}
\caption{The pressure as a function of the chemical potential of the holographic model for three different values of the parameter $\Lambda$. The dashed red line is the plot of the stiff nuclear EoS from Ref.~\cite{Hebeler:2013nza}.}
\label{Fig:PMu2}
\end{figure}

We also explored the behavior of phase transitions for a fixed stiff hadronic EoS. For this objective we have considered different values of the parameter $\Lambda$, such that the transition from the hadronic to the quark matter phases occurs at higher densities. The results are displayed in Fig.~\ref{Fig:PMu2}, where the EoS  of the holographic model is potted for three different values of the parameter $\Lambda$, namely, $125.965$ MeV (solid blue line), $133.372$ MeV (solid black line), and $162.475$ MeV (solid brown line). The stiff nuclear EoS from Ref.~\cite{Hebeler:2013nza} is also drawn in such a figure (dashed red line). The intersections are represented by black points and have, respectively, the coordinates $(\mu,\, p) = $ $(440.272\,\text{MeV},$ $ 103.679\,\text{MeV}/\text{fm}^3)$ for the case with $\Lambda=125.965\,\text{MeV}$ (solid blue line), $(\mu,\, p)=$ $(469.714\,\text{MeV},$ $  137.653\,\text{MeV}/\text{fm}^3)$ for the case with $\Lambda=133.373$ MeV (solid black line), and  $(\mu,\, p) =$ $(560.390\,\text{MeV},$ $ 260.259\,\text{MeV}/\text{fm}^3)$ for the case with  $\Lambda=162.475$ MeV (solid brown line). It can be seen that an increase in the parameter $\Lambda$ produces an increase in the transition pressure. 

\subsection{Stellar structure of a heavy quark compact stars}

This section is devoted to investigating the stellar structure of compact stars employing the equations of state discussed previously. In the sequence we choose geometrical units such that $G=1=c$. We consider the star as being composed of a perfect fluid whose energy-momentum tensor can be written in the form
\begin{equation}
{T}_{\mu\nu} =(\epsilon + p) u_{\mu}u_{\nu} + p{g}_{\mu\nu},
\end{equation}
with $\epsilon$, $p$, and $u_\mu$ being respectively the energy density, the pressure, and the four-velocity of the fluid. In addition, for the generic background space-time of a static spherical star, we use the following line element
\begin{equation}\label{dsz_tov}
ds^{2}=-e^{ \nu(r)} dt^{2} + e^{ \lambda(r)} dr^{2} + r^{2}(d\theta^{2}+\sin^{2}{\theta}d\phi^{2}) ,
\end{equation}
with $(t,\,r\,\,\theta,\, \varphi)$ being Schwarzschild-like coordinates, and $\nu(r)$ and $\lambda(r)$ are functions of the indicated coordinate alone. The Einstein equations in such a spacetime lead to the  well know Tolman-Oppenheimer-Volkov equations for stellar structure, 
\begin{eqnarray}
\label{tov1}
  &&\frac{dp}{dr} = - \frac{\epsilon m}{r^2}\bigg(1 + \frac{p}{\epsilon}\bigg)
	\bigg(1 + \frac{4\pi p r^3}{m}\bigg)\bigg(1 -
	\frac{2m}{r}\bigg)^{-1},\qquad
\\ \nonumber \\
\label{tov2}
\!\! &&\frac{d\nu}{dr} = - \frac{2}{\epsilon} \frac{dp}{dr}
	\bigg(1 + \frac{p}{\epsilon}\bigg)^{-1},
\\ \nonumber \\ 
\label{tov3}
\! \!&&\frac{dm}{dr} = 4 \pi r^2 \epsilon,
\end{eqnarray}
where $m$, $p$, $\epsilon$, and $\nu$ are function of the radius $r$ only. The integration of these equations requires initial conditions. As usual, we take $P(r=0)= p_c$ as the given central pressure, $m(r=0)=0$ and $\nu(r=0)=\nu_c$ as the given values for the mass and the metric field at the center of the star. During integration, we also use the EoS to obtain the energy density $\epsilon$ at each radius $r$ inside the star. In the cases of one phase hadronic stars, a unique equation of sate is used and the integration is performed from the center toward the surface. The integration process is finished when the pressure reaches the zero value. At this point we identify the coordinate $r = R$ as the star radius and  $M = m(R)$ as the star mass. In the cases of two-phases stars (hybrid stars) we use two equations of state and two separated numerical integration processes inside the star. The first one is the integration from the center to the interface (i.e., the place where the pressure reaches its transition value), and the second one is the integration from the interface to the surface of the star. 

We first investigate the structure for compact objects considering the holographic EoS for fixed $\Lambda=125.965\, {\rm Mev/fm^3}$, taking the input data for the central pressure from the situation of Fig.~\ref{Fig:PMu}. For the hybrid stars, the TOV equation is solved by taking the combination of two different equations of state, always considering the holographic EoS as a member used to describe the central region of the star.  The initial condition for each integration is the value of the central pressure, which depends on the specific parameters of the phase transition. In the case of the pair holographic/stiff EoSs, the central pressure values (in ${\rm Mev/fm^3}$) are in the interval $[2.14949,\, 16547.0]$, with the transition at $p_c= 107.013$, and the central energy density in the interval $[139.014,\, 53884.0]$, also in ${\rm MeV/fm^3}$. For central pressure values below $p_c= 107.013$, the hadronic (stiff) EoS is employed alone and the results are pure baryonic stars modeled by the stiff EoS. In the cases of the pair holographic/intermediate EoSs, the central pressure values are in the interval $[2.60088,\, 16547.0]$, with the transition at $p_c=  229.391$, and the central energy density in the interval $[177.027,\, 53884.0]$, also in ${\rm MeV/fm^3}$. For central pressure values below $p_c= 229.391$, the results are pure baryonic stars modeled by the intermediate EoS.
In the case of the pair holographic/soft EoSs, the central pressure values (in ${\rm Mev/fm^3}$) are in the interval $[2.60088,\, 16547.0]$, with the transition at $p_c= 594.982$, and the central energy density in the interval $[190.215,\, 53884.0]$, also in ${\rm MeV/fm^3}$. For central pressure values below $p_c=594.982$, the results are pure baryonic stars modeled by the soft EoS.

\begin{figure}[ht]
\centering
\includegraphics[width=8cm]{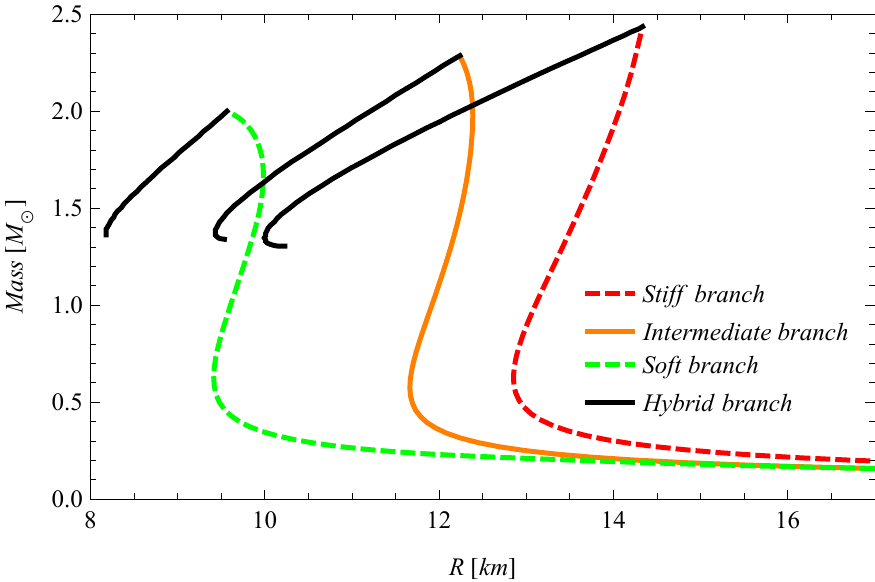}
\caption{
The figure shows the mass-radius relation of hybrid stars obtained by solving the TOV equations. The red dashed line represents the hadronic star branch corresponding to the stiff EoS, while the orange solid line to the intermediate EoS, the green dashed line to the soft EoS, and solid black lines represent hybrid stars build from the hadronic and the quark matter EoS (i.e. stars with phase transitions in its interior). These results were obtained for $\Lambda=125.965$ MeV and correspond to the results displayed in Fig~\ref{Fig:PMu}.}
\label{Fig:MR}
\end{figure}

The numerical results for the mass of the compact objects as a function of the respective radius are displayed in Fig.~\ref{Fig:MR}. We come up with three different stellar models composed of hadronic matter and quark matter, the so called hybrid stars, which correspond to the black solid lines in that figure. For pressure phase transition pure hadronic stars are obtained and they correspond to red dashed, orange, and green dashed lines, those stars were built respectively by using the stiff, intermediate, and soft EoS. As we can observe, this scenario supports stars with masses of $2.44$\(M_\odot\), $2.29$\(M_\odot\) and $2.00$\(M_\odot\), respectively. The corresponding radii are $14.3$ km, $12.2$ km, and $9.57$ km, respectively. It can be seen that these solutions are unstable under the static criterion of stability  \cite{Pereira:2017rmp} (see also \cite{Glendenning:1997wn}). It is also confirmed that more compact objects are the hybrid stars obtained with the combination of the holographic and the soft EoSs.

\begin{figure}[ht]
\centering
\includegraphics[width=8cm]{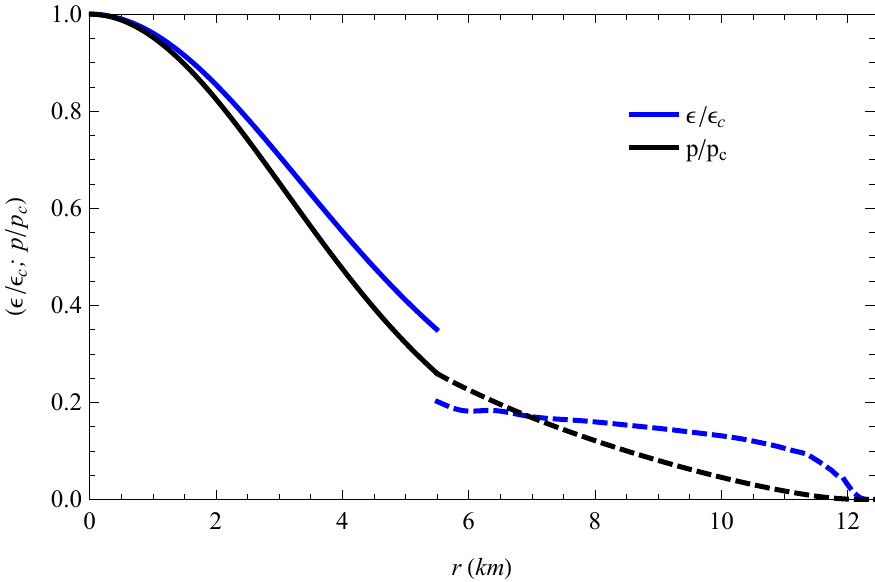}
\caption{The figure shows the evolution of the normalized energy density (blue) and normalized pressure (black) as a function of the radius in the interior of a single star. The quark matter EoS is represented by solid lines, while the stiff nuclear EoS by dashed lines. The central pressure is $p_c=400.000$ MeV/fm$^3$ and $\Lambda=125.965$ MeV.}
\label{Fig:MR3}
\end{figure}

In order to see some details of the internal structure of the objects modeled with the equations of state considered in the present study, we analyze a particular hybrid compact object. Figure~\ref{Fig:MR3} shows the evolution of the normalized energy density (in terms of the central density $\epsilon_c$) and of the normalized pressure (in terms of the central pressure $p_c$) described by the quark matter (solid line) and nuclear stiff (dotted line) EoSs across the chosen hybrid star. The central pressure and central density for this case are, respectively, $p_c= 400.000\, {\rm MeV/fm^3}$ and $\epsilon_c = 1862.48\, {\rm MeV/fm^3}$. As it can be seen, the energy density is discontinuous at the interface between the two matter phases.  The pressure is continuous but its derivative is not defined at the interface, which is located at the radial coordinate $r= 5.49295$ km, and the transition pressure is $p=1.04241\times 10^{2}\, {\rm MeV/fm^3}$.

\begin{figure}[ht]
\centering
\includegraphics[width=8cm]{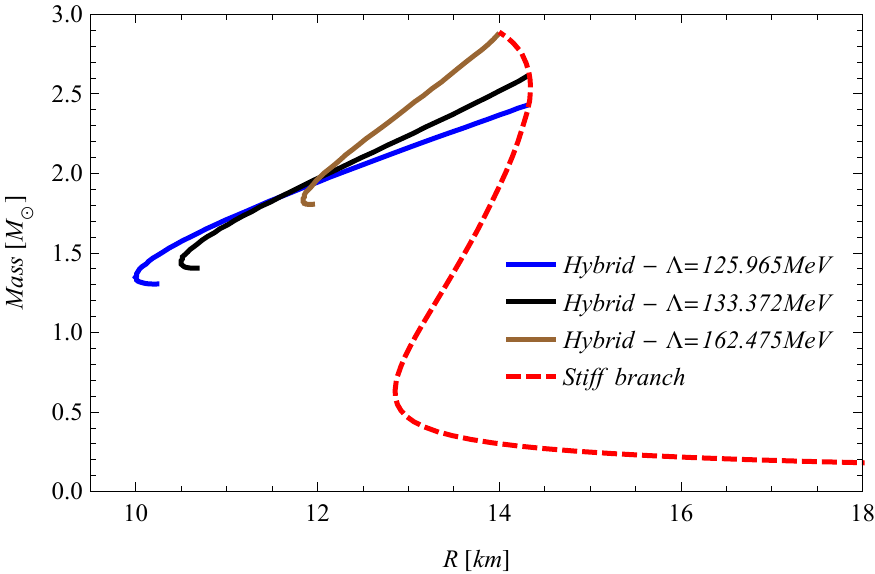}
\caption{
The figure shows the mass-radius relation of hybrid stars obtained by solving the TOV equations. The red dashed line represents the hadronic branch corresponding to the stiff EoS, while solid lines represent the hybrid branch build from the hadronic and quark matter EoS. These results were obtained for different values of $\Lambda$ and correspond to the results displayed in
Fig.~\ref{Fig:PMu2}}
\label{Fig:MR2}
\end{figure}

Finally, we solve the TOV equations for considering the holographic EoS for different values of $\Lambda$ matched to the stiff EoS model for the nuclear matter as displayed in Fig.~\ref{Fig:PMu2}. The results for the mass-radius relation of hybrid stars are displayed in Fig.~\ref{Fig:MR2}. We observe that the maximum mass depends on the transition point, the larger the chemical potential (and pressure) the larger the mass of the star. Thus, the maximum masses are: $2.42$\(M_\odot\), $2.61$\(M_\odot\) and $2.87$\(M_\odot\) for $\Lambda=125.965$ MeV, $\Lambda=133.372$ MeV and $\Lambda=162.475$ MeV, respectively. It is worth pointing out that these results are qualitatively equivalent to the results presented in Ref.~\cite{Fadafa:2019euu}. 

\section{Conclusion and outlook}
\label{Sec:Conclusions}

The thermodynamics and the phase structure of heavy quarks systems were studied in this work by using the holographic description. The interpretation of our results is motivated by a previous holographic model investigated in \cite{He:2013qq}, where the phase diagram is in agreement with a system composed by heavy quarks when contrasted with lattice QCD results. The holographic model used in this work was implemented in the context of Einstein-Maxwell-Dilaton theories. Its dual quantum field theory is at finite temperature and density. We show that the thermodynamic properties are sensitive to the value of chemical potential. 

We also found a first-order phase transition for $\mu<\mu_c$ and also showed that the critical point lies at $\mu=\mu_c$, where the phase transition becomes second-order. In turn, for $\mu>\mu_c$ the transitions become crossover. The phase diagram summarises these findings.

In order to investigate the inner structure of compact objects, we built the EoS for the heavy quark system. We realized that the free energy naturally decouples, allowing us to get an expression which is independent of the temperature, what make it possible to calculate the EoS. These findings allowed us to investigate the inner structure of stars composed by heavy quarks central core matched to an outer nuclear matter crust. Analogously to what was done in Ref.~\cite{Hoyos:2016zke} (see also \cite{Annala:2017tqz}), we explored the transition between the nuclear matter EoSs to quark matter EoS. It is worth mentioning that the EoS we got is qualitatively equivalent to the one obtained in the holographic top-down approach investigated in \cite{Hoyos:2016zke, Annala:2017tqz}. Indeed, the results of the pressure as a function of the chemical potential are quantitatively equivalent, showing the consistency of the top-down and bottom-up approaches for describing QCD-like theories. We conclude that in all cases the quark matter stars we found are unstable.

Finally, we stress that the present study of a bottom-up holographic model is the first step and we plan to expand it to try to find more realistic equations of state, what could be useful in order to describe the inner structure of compact objects, resulting in important applications in Astrophysics. We are also planning to address further studies, turning on the kinetic function, which allows us to get a phase diagram in agreement with the one expected in QCD with light quarks \cite{Yang:2014bqa}. Other directions of investigation also include the use of the holographic dictionary to include magnetic field \cite{He:2020fdi} and nucleons in the theory to model the matter inside compact objects.

{\bf Note added:} After the first version of this draft was announced in the arXiv, we saw the paper of Ref.~\cite{Ballon-Bayona:2020xls}, where the authors investigate the thermodynamics of heavy quarks within a model in the context of holographic EMD models for QCD. Their approach is quite different from ours. In particular, they use the dilaton field as input, while we use the warped factor instead. In both models, the thermodynamic analysis shows qualitatively equivalent results as well as the phase diagram.

\section*{Acknowledgments}

We would like to thank Jorge Noronha, Romulo Rougemont, Alfonso Ballon Bayona and Diego Rodrigues for stimulating discussions along the development of this work. L. A. H. M. and V. T. Z. thank financial support from Coordena\c{c}\~ao de Aperfei\c{c}oamento do Pessoal 
de N\'ivel Superior (CAPES, Brazil), Programa Nacional de P\'os-Doutorado, and 
 Grant No.~88881.310352/2018-01.
V. T. Z. also thanks financial support from Conselho 
Nacional de Desenvolvimento Cient\'\i fico
e Tecnol\'ogico (CNPq, Brazil), Grant No.~309609/2018-6.

\appendix

\section{Equation of state in AdS Einstein-Maxwell solution}
\label{AppendixA}

In this section, we investigate the extremal solution of the Einstein-Maxwell equations. In this case, the solution of the gauge field is given by
\begin{equation}\label{Eq:RNGaugeField}
A_0(z)=\mu-Q z^2.
\end{equation}
The condition at the horizon gives us $Q=\mu/z_h^2$. It is worth mentioning that the relation \eqref{Eq:RelationQmu} reduces to this result when $\Lambda$ goes to zero. While the horizon function solution is given by Eq.~\eqref{Eq:RNSol}. Then, the black hole temperature is 
\begin{equation}\label{Eq:RNTemperature}
T=\frac{1}{z_h\pi}\left(1-\frac{q^2 z_h^6}{2}\right).
\end{equation}
We also may calculate the entropy density using the relation \eqref{Eq:Entropy}
\begin{equation}\label{RNEntropy}
s=\frac{2\pi\ell^3}{\kappa^2\,z_h^3}.
\end{equation}
In the same way the free energy through \eqref{Eq:FreeEnergy}
\begin{equation}\label{Eq:RNFreeEnergy}
\mathcal{F}=-\frac{\ell^2}{2\kappa^2z_h^4}-\frac{5\ell\mu^2}{6\kappa^2z_h^2}.
\end{equation}
Analogously, the pressure is $p=-\mathcal{F}$
\begin{equation}\label{Eq:RNPressure}
p=\frac{\ell^2}{2\kappa^2z_h^4}+\frac{5\ell\mu^2}{6\kappa^2z_h^2}.
\end{equation}
Let us consider the relationship between the charge and the chemical potential, which is obtained plugging evaluating Eq. \eqref{Eq:RNGaugeField} at the horizon, then, plugging $Q=\mu/z_h^2$ in Eq. \eqref{Eq:RelationQq} we get
\begin{equation}
q^2=\frac{\mu^2}{3 z_h^4 \ell^2}.
\end{equation}
Then, we may write the temperature \eqref{Eq:RNTemperature} as a function of the chemical potential
\begin{equation}
T=\frac{1}{\pi z_h}\left(1-\frac{z_h^2 \mu^2}{6 \ell^2}\right).
\end{equation}
The last equation is a quadratic equation on $z_h$. Solving this equation we get a relation of $z_h$ as a function of the temperature and chemical potential
\begin{equation}\label{Eq:RNzhT}
z_h=\frac{-3\pi T \ell^2+\ell\left(9\pi^2 T^2 \ell^2+6 \mu^2\right)^{1/2}}{\mu^2}.
\end{equation}
Expanding the square root in the high temperature regime, keeping up to the first subleading term we get
\begin{equation}\label{Eq:RNzhApproximation}
z_h=\frac{1}{\pi T}-\frac{\mu^2}{6 \pi^3 \ell^2 T^3}+\mathcal{O}(1/T^5)
\end{equation}
Plugging \eqref{Eq:RNzhApproximation} and $\kappa$ from \eqref{Eq:RNTemperature} in \eqref{RNEntropy}, then, we expand in the high temperature regime, the asymptotic expression for the entropy becomes
\begin{equation}
s=\frac{2\pi^4\ell^3 T^3}{\kappa^2}+\frac{\pi^2\ell\mu^2 T}{\kappa^2}+\mathcal{O}\left(\mu^4/T,\mu^6/T^3\right)
\end{equation}
we observe that the entropy scales correctly with the temperature. In turn, plugging \eqref{Eq:RNzhT} in \eqref{Eq:RNFreeEnergy} the free energy is given by
\begin{equation}\label{Eq:RNFreeEnergyApprox}
\mathcal{F}=-\frac{\pi^4\ell^3 T^4}{2\kappa^2}-\frac{7\pi^2\ell \mu^2 T^2}{6\kappa^2}-\frac{11\mu^4}{36\ell\kappa^2}+\mathcal{O}\left(\mu^6/T^2\right),
\end{equation}
where $\mathcal{O}\left(\mu^6/T^2\right)$ contains higher-order contributions. Observing Eq. \eqref{Eq:RNFreeEnergyApprox} the expression decouples into one piece depending only on the temperature, chemical potential, and a mixture of the form $\mathcal{O}\left(\mu^6/T^2\right)$. The pressure is given by $p=-\mathcal{F}$, therefore, the pressure also decouples. Considering the piece depending only on the chemical potential, because we want to describe a cold phase in the dual field theory, we get
\begin{equation}
p=\frac{11\mu^4}{36\ell\kappa^2},
\end{equation}
whereas the energy density is given by
\begin{equation}
\epsilon=\frac{11\mu^4}{12\ell\kappa^2}.
\end{equation}
Hence, the EoS is $\epsilon=3p$, which is the EoS of a system preserving conformal symmetry.

On the other hand, let us investigate the extremal solution of the Reissner-Nordstrom AdS solution, which is given when the temperature \eqref{Eq:RNTemperature} vanishes, it happens when $q=\sqrt{2}/z_h^3$. It is worth mentioning that the location $z_h$ can be chosen close to the boundary, this means that the value of the black hole charge should be large. Plugging $Q=\mu/z_h^2$ and $q=\sqrt{2}/z_h^3$ in \eqref{Eq:RelationQq} we get a relation between $\mu$ and $z_h$
\begin{equation}
z_h=\frac{\sqrt{6}\ell}{\mu}.
\end{equation}
Then, plugging $q$ and $z_h$ in \eqref{RNEntropy}, \eqref{RNEntropy} and \eqref{Eq:RNPressure} we get
\begin{equation}\label{Eq:ExtremalAdSRNSol}
s=\frac{\pi\mu^3}{3\sqrt{6}\kappa^2},\quad \mathcal{F}=-\frac{\mu^4}{12\kappa^2\ell},\quad p=\frac{\mu^4}{12\kappa^2\ell}.
\end{equation}
Now, we may calculate the energy density using the relation $\epsilon=\mu\,\partial_{\mu}p-p$, which gives us
\begin{equation}
\epsilon=\frac{\mu^4}{4\kappa^2\ell}.
\end{equation}
Finally, the EoS of the extremal solution is
\begin{equation}\label{Eq:RNEoS}
\epsilon=3p,
\end{equation}
which is the EoS of a system preserving conformal symmetry. It is worth mentioning that in the extremal solution we did not do any approximation.

Finally, let us write the metric in the extremal case and expand it close to the horizon. Plugging $q=\sqrt{2}/z_h^3$ in \eqref{Eq:RNSol}
\begin{equation}
f(z)=(z^2-z^2_h)^2\frac{(2z^2+z_h^2)}{zh^6},
\end{equation}
then,
\begin{equation}
f(\tilde{z})\approx12\frac{\tilde{z}^{\,2}}{z_h^2}+28\frac{\tilde{z}^{\,3}}{z_h^3}+\mathcal{O}(\tilde{z}^{\,4}/z_h^4),
\end{equation}
where we have defined $\tilde{z}=z-z_h$. Plugging this result in \eqref{Eq:AnsatzMetric} and replacing $A=-\ln{(z/\ell)}$
\begin{equation}
ds^2=-12\frac{\ell^2\tilde{z}^{\,2}}{z_h^4}dt^2+\frac{\ell^2}{12\tilde{z}^{\,2}}d\tilde{z}^2+\frac{\ell^2}{z_h^2}dx_idx^i.
\end{equation}
Defining the variable $\tilde{z}=\hat{r}z_h^2/\ell^2$, thus, the metric becomes
\begin{equation}\label{Eq:RNExtremalMetric}
ds^2=-12\frac{\hat{r}^2}{\ell^2}dt^2+\frac{\ell^2}{12\hat{r}^2}d{\hat{r}}^2+\frac{\ell^2}{z_h^2}dx_idx^i.
\end{equation}
This is the $AdS_{2}\times \mathbb{R}^{3}$ metric, with boundary at $\hat{r}\to \infty$ and radius $\tilde{\ell}=\ell/\sqrt{12}$. The emergence of this geometry in the IR is related to the emergence of quantum criticality arising naturally in the holographic model, for details see Ref.~\cite{Faulkner:2009wj}.

\end{document}